\documentstyle[multicol,aps,epsf]{revtex}
\newcommand{\fr}{\frac}

\newcommand{\Ga}{\Gamma}
\newcommand{\al}{\alpha}
\newcommand{\be}{\beta}

\begin{document}

\title{A Deduced Feynman Rule for Calculating Retarded
     and Advanced Green Function\\
     in Closed Time Path Formalism}

\author{Jun Xiao and Enke Wang}

\address{Institute of Particle Physics, Huazhong Normal University,
Wuhan 430079}

\date{\today}

\maketitle

\begin{abstract}
Based on the closed time path formalism, a new Feynman rule for
directly calculating the retarded and advanced Green functions is
deduced. This Feynman rule is used to calculate the two-point
self-energy and three-point vertex correction in $\phi^3$ theory.
The generalized fluctuation-dissipation theorem for three-point
nonlinear response function is verified.


\end{abstract}

\begin{multicols}{2}

The closed time path formalism (CTPF)
\cite{Schwinger,Keldysh,Chou} in real-time finite-temperature
field theory has been used widely to investigate the equilibrium
and non-equilibrium properties
\cite{Wang1,Wang2,Hou,Aarts,Eijck,Aurenche} of the thermal system
recently. It is different to the imaginary time formalism (ITF),
without analytical continuation the physical region can be reached
in the real time formalism. In the CTPF two kinds of fields are
defined according to the closed time integral contour in the
generating function of Green function. The closed time integral
contour consists of the first branch which runs from negative
infinity to positive infinity and the second branch which runs
back from positive infinity to negative infinity. The field
located on the first and the second branch is defined as physical
(type-1) and ghost (type-2) field, respectively. So that the
propagator becomes a $2\times 2$ matrix with components $G_{11},
G_{12}, G_{21}, G_{22}$, their corresponding self-energy are
denoted as $\Sigma_{11}, \Sigma_{12}, \Sigma_{21}, \Sigma_{22}$
according to the type-1 or type-2 external leg of the Feynman
diagram. It is known that the self-energy with physical
interpretation is not $\Sigma_{11}$ but the retarded function
$\Sigma^{ret}$ (or advanced function $\Sigma^{adv}$) which is the
analytical continuation of the self-energy obtained in the
ITF\cite{Aurenche}. In order to calculate retarded (or advanced)
self-energy in the real time formalism, the usual way is to
calculate $\Sigma_{11}$, $\Sigma_{12}$, $\Sigma_{21}$ and
$\Sigma_{22}$ first, and then use the relation $\Sigma^{ret} =
\Sigma_{11} + \Sigma_{12}$ (or $\Sigma^{adv} = \Sigma_{11} +
\Sigma_{21}$) to get them. When one calculates n-point ($n>2$)
retarded/advanced vertex\cite{Wang3}, for example, three-point
fully retarded vertex $\Gamma_{raa}$ defined in
Ref.\cite{Lehmann}, we need to calculate eight components
$\Gamma_{111}$, $\Gamma_{112}$, $\Gamma_{121}$, $\Gamma_{122}$,
$\Gamma_{211}$, $\Gamma_{212}$, $\Gamma_{221}$ and $\Gamma_{222}$
first and then use the relation $\Gamma_{raa} = {1\over
2}(\Gamma_{111} - \Gamma_{112} - \Gamma_{121} + \Gamma_{122} +
\Gamma_{211} - \Gamma_{212} - \Gamma_{221} + \Gamma_{222})$ to get
it, the calculation becomes somewhat tedious. In this letter we
deduce a novel Feynman rule which can be used to calculate n-point
retarded/advanced Green function directly to avoid above tedious
calculation, and then we verify the generalized
Fluctuation-dissipation theorem for three-point nonlinear response
function by calculating the vertex correction.

In the CTPF, for any field $\phi$ the four components of the
matrix propagator in the single-time representation are defined as
 \begin{equation}
 \label{Ga1a2}
   G_{a_1 a_2}(x_1,x_2) \equiv -i
   \langle T_p\bigl(\phi_{a_1}(x_1)\phi_{a_2}(x_2)\bigr)\rangle \, ,
 \end{equation}
where $T_p$ represents the time ordering operator along the closed
time path, it is normal and  antichronological time ordering  on
the first and second branch of the closed time path, respectively.
$a_1,a_2{\,\in\,}\{1,2\}$ indicate on which of the two branches
the $\phi$ fields are located. $\langle\cdots\rangle$ stands for
the thermal expectation value. Following \cite{Chou} we define
 \begin{equation}
 \label{ar}
  \phi_a(x) = \phi_1(x){-}\phi_2(x) \, ,\quad
  \phi_r(x) = {\textstyle{1\over 2}} (\phi_1(x){+}\phi_2(x))\, ,
 \end{equation}
and the 2-point Green function in the above $(r,a)$ basis
 \begin{equation}
 \label{Gaf1af2}
  G_{\alpha_1\alpha_2}(x_1,x_2) \equiv -i 2^{n_r-1}
  \langle T_p\bigl(\phi_{\alpha_1}(x_1)\phi_{\alpha_2}(x_2)\bigr)\rangle ,
 \end{equation}
where $\alpha_1,\alpha_2{\,\in\,}\{a,r\}$, and $n_r$ is the number
of $r$ indices among $\{\alpha_1,\alpha_2\}$. Inserting
Eq.~(\ref{ar}) into (\ref{Gaf1af2}) it is easy to get
 \begin{eqnarray}
 G_{rr}(x_1, x_2) &=&G_{12}(x_1, x_2)+G_{21}(x_1,x_2)
   \nonumber\\
   &=&-i\langle[\phi(x_1),\phi(x_2)]_{\pm}\rangle\, ,
 \label{drr}\\
 G_{ra}(x_1, x_2) &=&G_{11}(x_1, x_2)-G_{12}(x_1, x_2)
 \nonumber\\
   &=&-i\theta(x_1^0-x_2^0)\langle [\phi(x_1),\phi(x_2)]_{\mp}\rangle\, ,
 \label{dra}\\
 G_{ar}(x_1, x_2) &=&G_{11}(x_1, x_2)-G_{21}(x_1, x_2)
   \nonumber\\
   &=&i\theta(x_2^0-x_1^0)\langle [\phi(x_1),\phi(x_2)]_{\mp}\rangle\, ,
  \label{dar}\\
 G_{aa}(x_1, x_2) &=& 0\, .
 \label{daa}
 \end{eqnarray}
Here
$[\phi(x_1),\phi(x_2)]_{\pm}=\phi(x_1)\phi(x_2)\pm\phi(x_2)\phi(x_1)$.
The double sign $\pm$ or $\mp$ in Eq.~(\ref{drr})-(\ref{dar})
corresponds to Boson and Fermion field for the upper and lower
case, respectively. Obviously $G_{ra}(x_1, x_2)$ and
$G_{ar}(x_1,x_2)$ are the usual retarded and advanced Green
functions.

Denote the $2\times 2$ matrix propagator in the single-time
representation as
\begin{equation}
\label{G}
  G=\left(
  \begin{array}{lcr}
  G_{11}  & G_{12} \\
  G_{21}  & G_{22}
  \end{array} \right) \, ,
\end{equation}
and the matrix propagator in the $(r,a)$ representation as
\begin{equation}
\label{Gbar}
  \overline{G}=\left(
  \begin{array}{lcr}
  G_{aa} & G_{ar} \\
  G_{ra} & G_{rr}
  \end{array} \right)
  =\left(
  \begin{array}{lcr}
  0 & G^{adv} \\
  D^{ret} & D^{col}
  \end{array} \right)\, .
\end{equation}
Both matrix propagators satisfy following transformation
relation\cite{Chou}:
 \begin{equation}
 \label{relation}
   G = Q^{\dag}\overline{G}Q\, ,
 \end{equation}
where $Q$ is orthogonal Keldysh transformation matrix
\begin{eqnarray}
  Q&=&\left(
  \begin{array}{lcr}
  Q_{a1} & Q_{a2} \\
  Q_{r1} & Q_{r2}
  \end{array} \right)
  =\frac{1}{\sqrt{2}}\left(
  \begin{array}{lcr}
  1 & -1 \\
  1 & 1
  \end{array} \right)\, ,
  \label{Q}\\
  Q^{\dag}&=&\left(
  \begin{array}{lcr}
  Q^{\dag}_{1a} & Q^{\dag}_{1r} \\
  Q^{\dag}_{2a} & Q^{\dag}_{2r}
  \end{array} \right)
  =\frac{1}{\sqrt{2}}\left(
  \begin{array}{lcr}
  1 & 1 \\
  -1 & 1
  \end{array} \right)\, .
  \label{Qdag}
\end{eqnarray}

\begin{figure}
 \epsfxsize 82mm \epsfbox{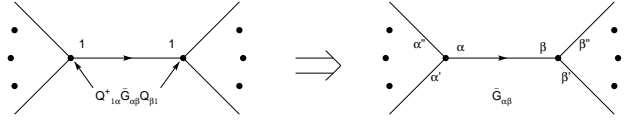}
 \vskip 0.2cm
 \caption{ The new bare vertex and propagator $\overline{G}_{\alpha\beta}$ after
 absorbing $Q^{\dag}$ and $Q$ into the left and right vertex
 in our deduced new Feynman rule.
 \label{fig1}}
 \end{figure}

\begin{figure}
 \epsfxsize 82mm\epsfbox{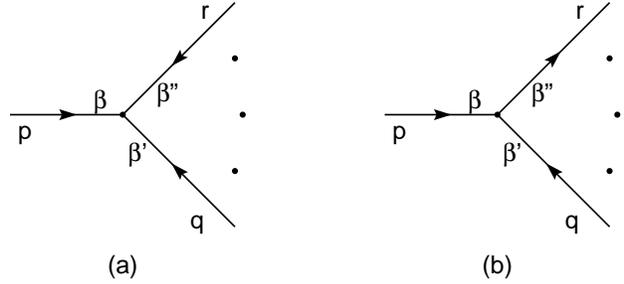}
 \vskip 0.2cm
 \caption{The bare vertex in the $(r,a)$ basis. The difference
 between (a) and (b) is that the orientation of momentum $r$ is
 changed.
 \label{fig2}}
 \end{figure}

From the above transformation (\ref{relation}) we see that, both
left and right side of the propagator $\overline{G}$ associate
with the transformation matrix $Q$. We can establish a new Feynman
rule in following way: as illustrated in Fig.~\ref{fig1}, the
matrices $Q^{\dag}$ can be absorbed into the left vertex with
outgoing momentum line, and $Q$ can be absorbed into the right
vertex with incoming momentum line, so that we leave
$\overline{G}_{\alpha\beta}$ in the $(r,a)$ basis as a new
propagator of the propagating line. After absorbing all $Q$ from
propagating line, the new bare vertex with all incoming momentum
(as illustrated in Fig.~\ref{fig2}(a)) can be defined as
\begin{equation}
\label{vertex}
  \gamma_{\beta\beta'\cdots\beta''}(p, q, \cdots r)=
  Q_{\beta a}Q_{\beta' b}\cdots Q_{\beta'' c}g_{ab\cdots c}\, ,
\end{equation}
where $a, b,\cdots, c\in 1, 2$ and $\beta,\beta',\cdots\beta''\in
a, r$, a summation over repeated indices is understood,
$p+q+\cdots+r=0$ because of energy-momentum conservation. In the
single-time representation, $g_{11\cdots 1}=-g_{22\cdots 2}=g$,
all other components vanish, here $g$ is the coupling constant. As
shown in Fig.~\ref{fig2}, we change the orientation of incoming
momentum $r$ in Fig.~\ref{fig2}(a) as outgoing line in
Fig.~\ref{fig2}(b) with energy-momentum conservation
$p+q+\cdots-r=0$. After absorbing $Q^{\dag}$ associated with
outgoing momentum $r$ and all $Q$ associated with incoming
momentum into the vertex, this new bare vertex in
Fig.~\ref{fig2}(b) should be
\begin{equation}
\label{vertex1}
  \gamma_{\beta\beta'\cdots\beta''}(p, q, \cdots -r)=
  Q_{\beta a}Q_{\beta' b}\cdots Q^{\dag}_{c \beta''}
  g_{ab\cdots c}\, .
\end{equation}
From Eq.(\ref{Q}) and (\ref{Qdag}) we know $Q_{\beta'' c}=
Q^{\dag}_{c \beta''}$. Insert it into Eq.(\ref{vertex1}) and then
compare to Eq.(\ref{vertex}), we see clearly that the bare vertex
$\gamma_{\beta\beta'\cdots\beta''}(p, q, \cdots r)$ with
$p+q+\cdots+r=0$ is the same as bare vertex
$\gamma_{\beta\beta'\cdots\beta''}(p, q, \cdots -r)$ with
$p+q+\cdots-r=0$. This property indicates that the new bare vertex
defined here is independent of the orientation of the momentum. In
the following we can drop the momentum arguments for the bare
vertex. Substituting $Q_{a1}, Q_{a2}, Q_{r1}, Q_{r2}$ in
Eq.(\ref{Q}) into Eq.(\ref{vertex}), we can express the n-point
new bare vertex in the $(r,a)$ basis as
\begin{equation}
\label{vertex2}
  \gamma_{\alpha_1\alpha_2\cdots\alpha_n}=({\frac{1}{\sqrt 2}})^n
  g[1-(-1)^{n_a(\alpha_1,\alpha_2,\cdots,\alpha_n)}]\, ,
\end{equation}
where $n_a(\alpha_1,\alpha_2,\cdots,\alpha_n)$ is the number of
$a$ indices among $\{\alpha_1,\alpha_2,\cdots,\alpha_n\}$. As this
number is even, the bare vertex is zero, so that in the $(r, a)$
basis there are a half of bare vertices vanish and other are only
related to the coupling constant and independent of the thermal
distribution function. This property will help us to simplify the
calculation greatly. For example, the explicit form of three-point
bare vertices read
\begin{eqnarray}
\label{3point}
  &&\gamma_{aaa}=\gamma_{arr}=\gamma_{rra}=\gamma_{rar}=\fr{g}{\sqrt 2}\, ,
  \nonumber\\
  &&\gamma_{aar}=\gamma_{ara}=\gamma_{rrr}=\gamma_{raa}=0\, .
\end{eqnarray}

 \begin{figure}
 \epsfxsize 75mm\epsfbox{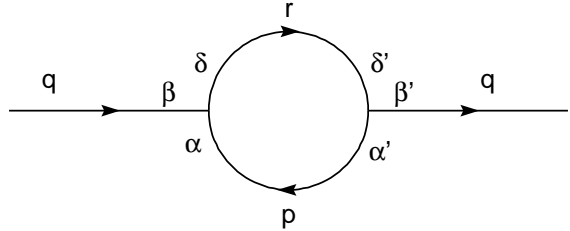}
 \vskip 0.2cm
 \caption{The one-loop Feynman diagram for calculating
          the retarded and advanced self-energy in $\phi^3$ theory.
 \label{fig3}}
 \end{figure}

As first application we use this new Feynman rule to calculate the
retarded (or advanced) self-energy in $\phi^3$ theory. As shown in
Fig.~\ref{fig3}, the self-energy in the $(r,a)$ basis can be
expressed as
\begin{eqnarray}
\label{self}
   -i\Sigma_{\beta\beta^{'}}(q) &=&
   \int{\frac{d^4p}{(2\pi)^4}}
   (-i\gamma_{\alpha\beta\delta})
 \nonumber\\
   &&\big(i\Delta_{\alpha\alpha^{'}}(p)\big)(-i\gamma_{\alpha^{'}\beta^{'}
   \delta^{'}})\big (\Delta_{\delta^{'}\delta}(r) \big)\, ,
\end{eqnarray}
where $r=p+q$. Noticing propagator $\Delta_{aa}=0$, the poles of
$\Delta_{ra}(p)\Delta_{ra}(r)$ and $\Delta_{ar}(p)\Delta_{ar}(r)$
are both on the same side of the real axis in the complex $p^0$
plane, we have
\begin{equation}
\label{p0integral}
   \int\limits_{-\infty}^{+\infty}{\frac{dp^0}{2\pi}}
   \Delta_{ra}(p)\Delta_{ra}(r)=
   \int\limits_{-\infty}^{+\infty}{\frac{dp^0}{2\pi}}
   \Delta_{ar}(p)\Delta_{ar}(r)=0\, .
\end{equation}
Then four components of the self-energy can be derived as
\begin{eqnarray}
   -i\Sigma_{ra}(q) &=& g^2\int\frac {d^4p}{(2\pi)^4}
   \{n(r_0)\Delta_{ar}(p)[\Delta_{ra}(r)-\Delta_{ar}(r)]
 \nonumber\\
   &&+ n(p_0)\Delta_{ra}(r)[\Delta_{ra}(p)-\Delta_{ar}(p)]\}\, ,
 \label{self1}\\
   -i\Sigma_{ar}(q) &=& g^2\int\frac {d^4p}{(2\pi)^4}
   \Big\{n(p_0)\Delta_{ar}(r)[\Delta_{ra}(p)-\Delta_{ar}(p)]
 \nonumber\\
   &&+ n(r_0)\Delta_{ra}(p)[\Delta_{ra}(r)-\Delta_{ar}(r)]\Big\}\, ,
 \label{self2}\\
   -i\Sigma_{aa}(q)&=&[2n(q_0)+1][\Sigma_{ra}(q)-\Sigma_{ar}(q)]\, ,
 \label{self3}\\
   -i\Sigma_{rr}(q)&=&0\, .
\label{self4}
\end{eqnarray}
Here $n(k_0)=1/[\exp(\beta k_0)-1]$ is the Bose thermal
distribution. It is easy to show that $\Sigma_{ar}(q)$ and
$\Sigma_{ra}(q)$ are the retarded and advanced self-energy which
correspond to the retarded and advanced analytical continuation of
the self-energy calculated in the ITF\cite{Weldon}.
Eq.(\ref{self3}) corresponds to the Fluctuation-Dissipation
theorem in linear response theory\cite{Callen}.

\begin{figure}
 \epsfxsize 65mm\epsfbox{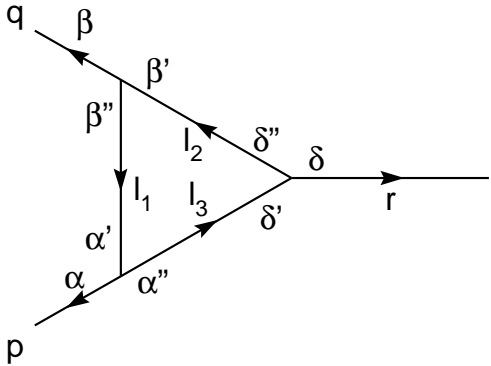}
 \vskip 0.2cm
 \caption{The Feynman diagram for calculating one-loop vertex
          correction in the $(r,a)$ basis in $\phi^3$ theory.
 \label{fig4}}
 \end{figure}

The one-loop three-point vertex correction is illustrated in
Fig.~\ref{fig4}, the general expression for this vertex correction
in the $(r,a)$ basis can be expressed as
\begin{eqnarray}
\label{three-point}
   -i\Gamma_{\al\be\delta}(p, q, r)&=& \int
   {\fr {d^4l_1}{(2\pi)^4}}(-i\gamma_{\al\al^{'}\al^{''}})
   [i\Delta_{\al^{'}\be^{''}}(l_1)]
 \nonumber\\
   &&(-i\gamma_{\be\be^{'}\be^{''}})
   [i\Delta_{\beta^{'}\delta^{''}}(l_2)]
  \nonumber\\
   &&(-i\gamma_{\delta\delta^{'}\delta^{''}})
   [i\Delta_{\delta^{'}\alpha^{''}}(l_3)]\, ,
\end{eqnarray}
where $ p+q+r=0$ and $l_1+l_2+l_3=0$.

Using $\Delta_{aa}=0$ and the $l_1^0$-integral over terms of the
type $\Delta_{ra}(l_1)\Delta_{ra}(l_2)\Delta_{ra}(l_3)$ and
$\Delta_{ar}(l_1)\Delta_{ar}(l_2)\Delta_{ar}(l_3) $ vanish because
the poles of these two terms are both on the same side of the real
axis in the complex $l_1^0$ plane, we deduce eight components of
one-loop vertex correction in $\phi^3$ theory as
\begin{eqnarray}
    &&-i\Ga_{rra}(p,q,r)= \fr{2g^3}{(\sqrt 2)^3}\int\fr{d^4l_1}{(2\pi)^4}
  \nonumber\\
    &&\quad\Big \{ n(l_{01})[\Delta_{ra}(l_1)
      {-}\Delta_{ar}(l_1)]\Delta_{ar}(l_2)\Delta_{ra}(l_3)
  \nonumber\\
    &&\quad + n(l_{02})[\Delta_{ra}(l_2){-}\Delta_{ar}(l_2)]
    \Delta_{ra}(l_1)\Delta_{ra}(l_3)
  \nonumber\\
    &&\quad +n(l_{03})[\Delta_{ra}(l_3){-}\Delta_{ar}(l_3)]
    \Delta_{ar}(l_1)\Delta_{ar}(l_2)\Big\},
  \label{point1}\\
    &&-i\Ga_{rar}(p,q,r)= \fr{2g^3}{(\sqrt 2)^3}\int\fr{d^4l_1}{(2\pi)^4}
  \nonumber\\
    && \quad \Big\{n(l_{01})[\Delta_{ra}(l_1){-}\Delta_{ar}(l_1)]
    \Delta_{ra}(l_2)\Delta_{ra}(l_3)
  \nonumber\\
    && \quad+ n(l_{02})[\Delta_{ra}(l_2) {-}\Delta_{ar}(l_2)]
    \Delta_{ar}(l_1)\Delta_{ar}(l_3)
  \nonumber\\
    && \quad + n(l_{03})[\Delta_{ra}(l_3){-}\Delta_{ar}(l_3)]
    \Delta_{ar}(l_1)\Delta_{ra}(l_2)\Big\},
  \label{point2}\\
    && -i\Ga_{arr}(p,q,r)= \fr{2g^3}{(\sqrt 2)^3} \int\fr{d^4l_1}{(2\pi)^4}
  \nonumber\\
    && \quad \Big\{n(l_{01})[\Delta_{ra}(l_1){-}\Delta_{ar}(l_1)]
    \Delta_{ar}(l_2)\Delta_{ar}(l_3)
  \nonumber\\
    && \quad +n(l_{02})[\Delta_{ra}(l_2){-}\Delta_{ar}(l_2)]
    \Delta_{ra}(l_1)\Delta_{ar}(l_3)
  \nonumber\\
    && \quad +n(l_{03})[\Delta_{ra}(l_3){-}\Delta_{ar}(l_3)]
    \Delta_{ra}(l_1)\Delta_{ra}(l_2)\Big\},
  \label{point3}\\
    &&-i\Ga_{aar}(p,q,r)=
      \fr{g^3}{(\sqrt 2)^3}\int\fr{d^4l_1}{(2\pi)^4}
  \nonumber\\
    && \quad \Big \{ [\Delta_{ra}(l_1){-}\Delta_{ar}(l_1)]
    \Delta_{ra}(l_2)\Delta_{ar}(l_3)[N_2{-}N_3]N_1
  \nonumber\\
    && \quad +[\Delta_{ra}(l_1){-}\Delta_{ar}(l_1)]
    \Delta_{ar}(l_2)\Delta_{ar}(l_3)[1{-}N_1N_2]
  \nonumber\\
    &&\quad+[\Delta_{ra}(l_1){-}\Delta_{ar}(l_1)]
    \Delta_{ra}(l_2)\Delta_{ra}(l_3)
       [N_1N_3{-}1]\Big\},
  \label{point4}\\
    &&-i\Ga_{raa}(p,q,r)= \fr{g^3}{(\sqrt 2)^3}\int\fr{d^4l_1}{(2\pi)^4}
  \nonumber\\
    && \quad \Big \{[\Delta_{ra}(l_2){-}\Delta_{ar}(l_2)]
    \Delta_{ra}(l_1)\Delta_{ra}(l_3)[N_1N_2{-}1]
  \nonumber\\
    && \quad +[\Delta_{ra}(l_2){-}\Delta_{ar}(l_2)]
    \Delta_{ar}(l_1)\Delta_{ra}(l_3)
      [N_2N_3{-}N_2N_1]
  \nonumber\\
    &&\quad+[\Delta_{ra}(l_2){-}\Delta_{ar}(l_2)]
    \Delta_{ar}(l_1)\Delta_{ar}(l_3)[1{-}N_2N_3]\Big\},
  \label{point5}\\
    &&-i\Ga_{ara}(p,q,r)= \fr{g^3}{(\sqrt 2)^3}\int\fr{d^4l_1}{(2\pi)^4}
  \nonumber\\
     && \quad \Big \{[\Delta_{ra}(l_3){-}\Delta_{ar}(l_3)]
     \Delta_{ar}(l_1)\Delta_{ar}(l_2)[1{-}N_1N_3]
  \nonumber\\
     && \quad +[\Delta_{ra}(l_3){-}\Delta_{ar}(l_3)]
     \Delta_{ra}(l_1)\Delta_{ar}(l_2)
       [N_1N_3{-}N_2N_3]
  \nonumber\\
     &&\quad+[\Delta_{ra}(l_3){-}\Delta_{ar}(l_3)]
     \Delta_{ra}(l_1)\Delta_{ra}(l_2)[N_2N_3{-}1]\Big\},
  \label{point6}\\
     &&-i\Ga_{aaa}(p,q,r)= \fr{g^3}{(\sqrt 2)^3}\int\fr{d^4l_1}{(2\pi)^4}
        \Big\{N_1N_2N_3
  \nonumber\\
     &&\quad\times [\Delta_{ra}(l_1)\Delta_{ra}(l_2)\Delta_{ra}(l_3)
        {-}\Delta_{ar}(l_1)\Delta_{ar}(l_2)\Delta_{ar}(l_3)]
  \nonumber\\
     &&\quad +(N_1{-}N_1N_2N_3)[\Delta_{ra}(l_1){-}
     \Delta_{ar}(l_1)]\Delta_{ra}(l_2)\Delta_{ar}(l_3)
  \nonumber\\
     &&\quad +(N_2{-}N_1N_2N_3)[\Delta_{ra}(l_2){-}
     \Delta_{ar}(l_2)]\Delta_{ar}(l_1)\Delta_{ar}(l_3)
  \nonumber\\
     &&\quad
        +(N_3{-}N_1N_2N_3)[\Delta_{ra}(l_3){-}\Delta_{ar}(l_3)]
  \nonumber\\
     &&\quad\times\Delta_{ra}(l_1)\Delta_{ar}(l_2)\Big\},
  \label{point7}\\
     &&-i\Gamma_{rrr}(p,q,r)=0.
  \label{point8}
\end{eqnarray}
Here $N_i=1+2n(l_{0i})$. Eq.(\ref{self4}) and (\ref{point8})
indicate the two-point self-energy and three-point vertex with all
indices being $r$ vanish, this property is also true for n-point
vertex in $n>3$ case\cite{Wang3}.

Denote $N_{k^0}=1+2n(k^0)$ for $k^0=p^0, q^0, r^0$. As $p+q+r=0$
and $l_1+l_2+l_3=0$, we have
\begin{eqnarray}
    && N_{p^0}[N_3-N_1]=N_3 N_1 -1\, ,
  \label{identity1}\\
    &&N_{q^0}[N_1-N_2]=N_1 N_2 -1\, ,
  \label{identity2}\\
    && N_{r^0}[N_2-N_3]=N_2 N_3 -1\, .
  \label{identity13}
\end{eqnarray}
Using above equations we can verify that three-point vertex
corrections in Eq.(\ref{point1})-(\ref{point8}) satisfy following
relations:
\begin{eqnarray}
   \Gamma_{ara}&=&N_{p^0}(\Gamma^*_{rar}-\Gamma_{rra})+N_{r^0}(\Gamma^*_{rar}-
   \Gamma_{arr})\, ,
 \label{FDT1}\\
   \Gamma_{raa}&=&N_{q^0}(\Gamma^*_{arr}-\Gamma_{rra})+N_{r^0}(\Gamma^*_{arr}-
   \Gamma_{rar})\, ,
 \label{FDT2}\\
   \Gamma_{aar}&=&N_{p^0}(\Gamma^*_{rra}-\Gamma_{rar})+N_{q^0}(\Gamma^*_{rra}-
   \Gamma_{arr})\, ,
 \label{FDT3}\\
   \Gamma_{aaa}&=&\Gamma^*_{arr}+\Gamma^*_{rar}+\Gamma^*_{rra}+N_{q^0}N_{r^0}
   (\Gamma_{arr}+\Gamma^*_{arr})
 \nonumber\\
   &&+ N_{p^0}N_{r^0}(\Gamma_{rar}+\Gamma^*_{rar})
 \nonumber\\
   &&+ N_{p^0}N_{q^0}(\Gamma_{rra}+\Gamma^*_{rra})\, .
 \label{FDT4}
\end{eqnarray}
These relations are just the generalized Fluctuation-Dissipation
theorem for three-point nolinear response function which is in
agreement with the result obtained in our previous
work\cite{Wang3}. Eq.(\ref{FDT1})-(\ref{FDT4}) together with
Eq.(\ref{point8}) indicate that, among these vertex functions
$\{\Gamma_{rra},\Gamma_{rar},\Gamma_{arr},\Gamma_{raa},
\Gamma_{ara}, \Gamma_{aar},\Gamma_{aaa},\Gamma_{rrr}\}$ only three
components are independent.

In Ref.\cite{Aurenche} Aurenche and Becherrawy developed a new
Feynman rule which denoted by $R, A$ to calculate retarded and
advanced vertex function. The main difference between their and
our Feynman rule is following: In the $(R, A)$ Feynman rule the
matrix propagator is diagonal and independent of the thermal
distribution function; the thermal distribution function is
absorbed into the vertex, the defined new vertex depends on the
relative orientation of the momenta and only two of the vertex
components vanish. In our $(r, a)$ Feynman rule the matrix
propagator is non-diagonal, one component is zero and another
component $\Delta_{rr}$ is related to thermal distribution
function; the defined new bare vertex is independent of thermal
distribution function and only related to the coupling constant;
this new bare vertex doesn't rely on the orientation of the
momenta and one half of the bare vertex components vanish. So that
in the practical calculation both $(R, A)$ and $(r, a)$ Feynman
rule have their own advantage and disadvantage. We should point
out that, in our $(r, a)$ basis the Green function is defined with
time-ordering and have explicit physical interpretation (see
Eq.(\ref{drr})-(\ref{daa}) and Ref.\cite{Wang3}), it works in both
coordinate and momentum space; but the $(R,A)$ Feynman rule, as
stated by the authors themselves in Ref.\cite{Aurenche}, only
works in the momentum space, and the $(R, A)$ Green function have
no explicit reference to a definite time-ordering.

By suitable combination we can get the $(R,A)$ vertex from the
$(r, a)$ vertex. For three-point vertex the relation between them
can be expressed as
\begin{eqnarray}
    \tilde{\Gamma}_{RRR}&=&\tilde{\Gamma}_{AAA}=0\, ,
  \label{aur1}\\
    \tilde{\Gamma}_{RRA}&=&\Gamma_{rra}\, ,
  \label{aur2}\\
    \tilde{\Gamma}_{RAR}&=&\Gamma_{rar}\, ,
  \label{aur3}\\
    \tilde{\Gamma}_{ARR}&=&\Gamma_{arr}\, ,
  \label{aur4}\\
    \tilde{\Gamma}_{AAR}&=&-\fr{1}{2}[N_{p^0}+N_{q^0}] \Gamma_{rra}^{*}\, ,
  \label{aur5}\\
    \tilde{\Gamma}_{ARA}&=&-\fr{1}{2}[N_{p^0}+N_{r^0}]\Gamma_{rar}^{*}\, ,
  \label{aur6}\\
    \tilde{\Gamma}_{RAA}&=&-\fr{1}{2}[N_{p^0}+N_{r^0}]\Gamma_{arr}^{*}\, .
  \label{aur7}
\end{eqnarray}

In summary, the new Feynman rule in the $(r,a)$ basis is deduced
from the Keldysh transformation in the CTPF. This Feynman rule can
be used to calculate the retarded and advanced Green function
directly. In this new Feynman rule, the bare vertex is only
related with the coupling constant and independent of the
orientation of the momenta, one half of bare vertex components
vanish; the new propagator depends on retarded, advanced
propagator and the thermal distribution function. As application,
the one-loop self-energy and three-point vertex correction in
$\phi^3$ theory are calculated in the $(r,a)$ basis, and the
Fluctuation-Dissipation theorem is verified in the two-point and
three-point cases. The difference between our $(r,a)$ Feynman rule
and the $(R,A)$ Feynman rule introduced by Aurenche and
Becherrawy\cite{Aurenche} is discussed.

\acknowledgments

This work was supported by the National Natural Science Foundation
of China (NSFC) under Grant Nos. 10135030 and 19928511.


\end{multicols}
\end{document}